\documentclass[aps,manuscript]{revtex4}
\usepackage{graphics}
\begin{document}
\newcommand{\bstfile}{aps} %alternative styles: osa, prasty or revtex
\newcommand{\bibs}{d:/Dad/Bibliography/TexBiB/final}
\draft
\title{Maximizing the hyperpolarizability of one-dimensional systems}
\author{Urszula B. Szafruga}
\address{Department of Physics and Astronomy, Washington State University, Pullman, Washington  99164-2814}
\author{David S. Watkins}
\address{Department of Mathematics, Washington State University, Pullman, Washington  99164-3113}
\author{Mark G. Kuzyk}
\address{Department of Physics and Astronomy, Washington State University, Pullman, Washington  99164-2814}

\begin{abstract}

Previous studies have used numerical methods to optimize the hyperpolarizability of a one-dimensional quantum system.  These studies were used to suggest properties of one-dimensional organic molecules, such as the degree of modulation of conjugation, that could potentially be adjusted to improve the nonlinear-optical response.  However, there were no conditions set on the optimized potential energy function to ensure that the resulting energies were consistent with what is observed in real molecules.  Furthermore, the system was placed into a one-dimensional box with infinite walls, forcing the wavefunctions to vanish at the ends of the molecule.  In the present work, the walls are separated by a distance much larger than the molecule's length; and, the variations of the potential energy function are restricted to levels that are more typical of a real molecule.  In addition to being a more physically-reasonable model, our present approach better approximates the bound states and approximates the continuum states -  which are usually ignored.  We find that the same universal properties continue to be important for optimizing the nonlinear-optical response, though the details of the wavefunctions differ from previous result.

\end{abstract}

\maketitle

\section{Introduction}

The nonlinear-optical susceptibility is a material property that describes the strength of light-matter interactions and is the basis for applications such as optical switching, which is used in telecommunications,\cite{wang04.01}
three-dimensional nano-photolithography used in making small structures,\cite{cumps99.01,kawat01.01}
and making new materials\cite{karot04.01} for novel cancer
therapies.\cite{roy03.01}  Quantum calculations show
that there is a limit to the nonlinear-optical response.\cite{kuzyk00.02,kuzyk01.01,kuzyk00.01,kuzyk03.01,kuzyk03.02,kuzyk04.02}
This limit provides a target for making optimized materials and is useful for defining scaling laws that can be used to determine the intrinsic properties of a molecule.  In this work, we focus on
the second-order susceptibility and the underlying molecular
hyperpolarizability, which is the basis of electro-optic switches and
frequency doublers.

We consider linear molecules in a potential well.  In previous work,\cite{zhou06.01,zhou07.02}
the molecule was situated in an infinite well, and arbitrarily large
variations of the potential were allowed.  In the current work, in an
effort to make our molecules more realistic, we model them with
a potential well of a depth not exceeding 8 eV, and we place the walls
of the infinite well far from the molecule compared with its electronic size.  As described later, under these
conditions we were able to obtain intrinsic hyperpolarizabilities of
as much as 0.708, which is about as big as we got in the previous
studies.\cite{zhou06.01,zhou07.02}.

The fundamental limit of the off-resonance hyperpolarizability is
given by,\cite{kuzyk00.01}
\begin{equation}\label{limit}
\beta_{MAX} = \sqrt[4]{3} \left( \frac {e \hbar} {\sqrt{m}}
\right)^3 \cdot \frac {N^{3/2}} {E_{10}^{7/2}} ,
\end{equation}
where $N$ is the number of electrons and $E_{10}$ the energy
difference between the
first excited state and the ground state, $E_{10} = E_1 - E_0$.  Using
Equation \ref{limit}, we can define the off-resonant intrinsic
hyperpolarizability, $\beta_{int}$, as the ratio of the actual
hyperpolarizability (measured or calculated), $\beta$, to the
fundamental limit,\cite{zhou08.01}
\begin{equation}\label{intrinsic-beta}
\beta_{int} = \beta / \beta_{MAX} .
\end{equation}
The intrinsic hyperpolarizability is a scale-invariant quantity because it does no depend on the number of electrons or on scale, as defined in the literature.\cite{kuzyk10.01}  Thus, it allows one to compare molecules of very different structures and sizes.
We note that since the dispersion of the fundamental limit of $\beta$
is also known,\cite{kuzyk06.03} it is  possible to calculate the
intrinsic hyperpolarizability at any set of wavelengths for any
second-order phenomena.  In the present work, we treat only the
zero-frequency limit.

Prior to 2007, an analysis of a large set of molecules showed that the largest nonlinear susceptibilities of the best
ones fell short of the of fundamental limit by a factor of about 30\cite{kuzyk03.02,Kuzyk03.05,Tripa04.01}, or $\beta_{int} \leq 0.03$.  This shortfall was shown to not be of a fundamental nature.\cite{Tripa04.01} Later, a molecule with asymmetric conjugation of modulation was measured to have $\beta_{int} = 0.048$,\cite{perez07.01} suggesting that even larger values might be possible.

In the present work, we apply numerical optimization using methods similar to that of Zhou and coworkers.\cite{zhou06.01}  This work led Zhou and coworkers to propose that modulated conjugation in the bridge between donor and acceptor ends of a molecule may be a new paradigm for making molecules with higher intrinsic hyperpolarizability,\cite{zhou06.01} a hypothesis that was experimentally investigated by P\'{e}rez Moreno.\cite{perez07.01}  Here, we investigate weather or not the same behavior is observed in our more restricted parameter space.

We also investigate universal scaling, the observation that a broad range of quantum systems whose hyperpolarizability is at the fundamental limit share certain properties.

\section{Computational Approach}

Each one-dimensional molecule is modeled by a potential function.
The potential is fixed at zero in the buffer
regions and takes on negative values between 0 and -8 eV in the region that
represents the molecule.  In that region the potential function is
piecewise linear with 39 degrees of freedom.
In previous studies we
have used cubic splines to represent the potential functions.  We
switched to piecewise polynomials so that the potential would nowhere
inadvertently overshoot the constraints $0 > V(x) > -8 \, eV$.

Starting from a given potential function, we use the Nelder-Mead
simplex algorithm\cite{lagar98.01} to vary the potential
to maximize $\beta_{int}$.  Since there are 39 degrees of freedom, we
are maximizing over a 39-dimensional space.
We have three ways of computing $\beta$,
all of which require solving the one-dimensional Schroedinger
eigenvalue problem for the given potential (and in some cases also for
neighboring potentials).  We solve the eigenvalue problem  numerically
on a computational mesh
consisting of 400 quadratic finite elements\cite{zienk05.01}
with a total of 799 degrees of freedom.  Half of the elements are
devoted to the part of
the computational domain that represents the molecule, and the other
half cover the buffer regions between the molecule and the infinite
walls.  The mesh is finest in the region that
represents the molecule and becomes coarser as one moves from the
molecule toward either wall.

Once we have solved the eigenvalue problem, we can compute transition
moments and then obtain $\beta$ by
the standard Orr and Ward SOS expression
$\beta_{SOS},$\cite{orr71.01}  the dipole free expression
$\beta_{DF}$,\cite{kuzyk05.02} or a finite difference approximation
$\beta_{NP}$, which is described in the literature.\cite{zhou07.01,zhou07.02}  In the optimization code we
use $\beta_{NP}$.  That is, we seek to maximize $\beta_{int} =
\beta_{NP}/\beta_{MAX}$.  Once the optimization is complete, we use
$\beta_{SOS}$ and $\beta_{DF}$ for comparison to check the accuracy of the result.

The exact computation of $\beta_{SOS}$ and $\beta_{DF}$
requires sums over infinitely many states.
We approximate them by summing over the 80 lowest energy
levels.   This is overkill; typically 20 or 30 states give a
sufficiently accurate approximation.  All computations are done using MATLAB.

In addition to calculating the hyperpolarizability using the three equivalent methods, we also
compute the matrix $\tau$, which represents deviations from the sum rules and is defined by\cite{zhou06.01,kuzyk06.01}
\begin{equation}
\tau_{mp}^{(N)} = \delta_{m,p} - \frac {1} {2} \sum_{n=0}^{N} \left(
 \frac {E_{nm}} {E_{10}} + \frac {E_{np}} {E_{10}}\right) \frac
 {x_{mn}} {x_{10}^{max}} \cdot \frac {x_{np}} {x_{10}^{max}} ,
 \label{tau}
\end{equation}
where $x_{10}^{max}$ is the magnitude of the fundamental limit of the
position matrix element $x_{10}$ for a one electron system, and is
given by,\cite{kuzyk00.01}
\begin{equation}
x_{10}^{max} = \frac {\hbar} {\sqrt{2m E_{10}}} . \label{x10MAX}
\end{equation}
Each matrix element of $\tau^{(N)}$, indexed by $m$ and $p$, is a
measure of how well the $(m,p)$ sum rule is obeyed when truncated to
$N$ states.  If the sum rules are exactly obeyed,
$\tau_{mp}^{(\infty)}=0$ for all $m$ and $p$.  We use 80 states (N = 80)
when calculating the $\tau$ matrix or the hyperpolarizability with an
SOS expression so that truncation errors are kept to a minimum.  In addition, since
the hyperpolarizability depends critically on the transition dipole
moment from the ground state to the excited states, we use the value
of $\tau_{00}^{(80)}$ as an important test of the accuracy of the
calculated wavefunctions.

\section{Results and Discussions}

Since the Nelder-Mead algorithm only gives a local optimum, we
arrive at different optimized potentials from different starting potentials.
We tried the same starting potential functions as used by Zhou and coworkers
\cite{zhou07.02}.  More precisely, the shapes
were the same, but we rescaled them to fit
within our new constraints.

\begin{figure}
\includegraphics{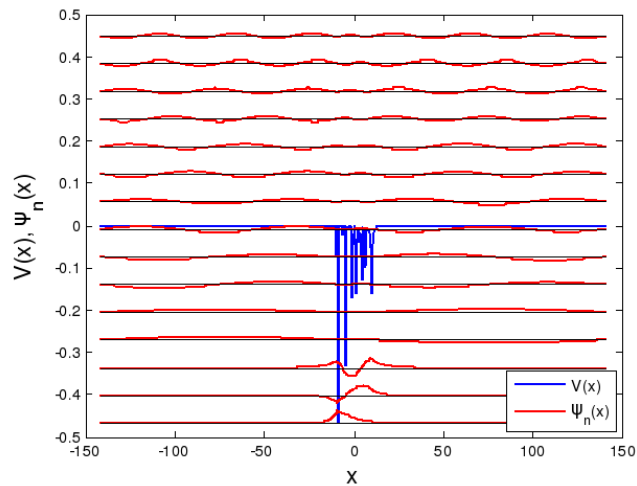}
\caption{Optimized potential energy function and first 15
  wavefunctions after 9450 iterations.  Starting potential is
  $V(x)=0$.}
\label{fig:V=0-Potential-wavefunctions}
\end{figure}

\begin{figure}
\includegraphics{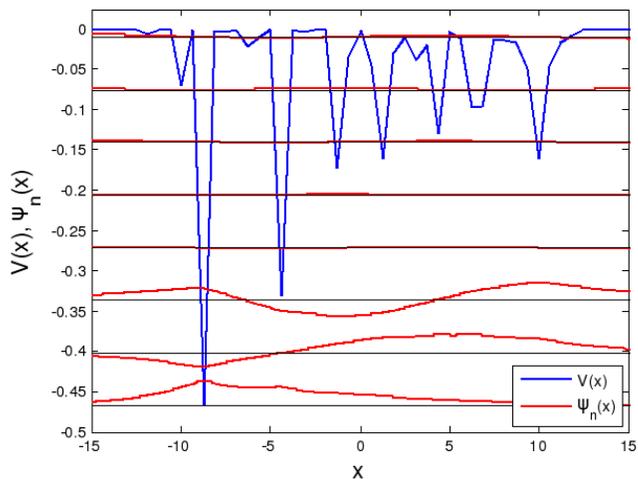}
\caption{Close up view of
Figure~\ref{fig:V=0-Potential-wavefunctions} showing the part of the domain
  that represents the molecule.}
\label{fig:V=0-close-up}
\end{figure}

Figure \ref{fig:V=0-Potential-wavefunctions} shows an example of the
optimized potential energy function after 9,450 iterations when
starting with the potential $V(x) = 0$.  Figure \ref{fig:V=0-close-up} shows an expanded view of only the potential well of the molecule.  Also shown in Figures \ref{fig:V=0-Potential-wavefunctions} and \ref{fig:V=0-close-up} are the eigenfunctions of the first 15 states and 8 states, respectively, computed from the
optimized potential.  First, we note that the potential energy
function shows the same kinds of wiggles as in our original
paper,\cite{zhou06.01} though not of sufficient amplitude to localize
the wavefunctions.

\begin{figure}
\includegraphics{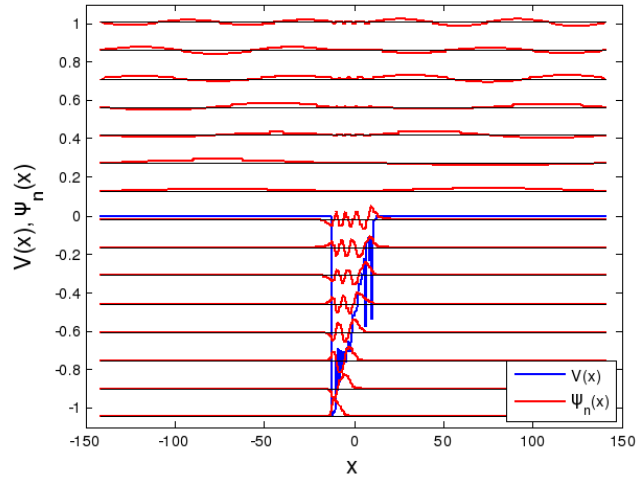}
\caption{Optimized potential energy function and first 15
  wavefunctions after 4,079 iterations.  Starting potential is $V(x)= x$.}
\label{fig:vxx}
\end{figure}

In previous work,\cite{zhou07.02} Zhou and coworkers found that the intrinsic hyperpolarizability is optimized for two broad classes of potential energy functions.  One in which the potential energy function is characterized by wiggles, as is the type shown in Figure \ref{fig:V=0-Potential-wavefunctions}, and another, in which the potential energy functions are relatively smooth, as shown in Figure \ref{fig:vxx}, an example of the optimized potential energy function when starting with the potential $V(x) = x$.  (Also shown are the eigenfunctions of the first 15 states computed with
the optimized potential.)  However, in contrast to past work, both of these potentials lead to a large degree in overlap between the energy eigenfunctions.

\begin{figure}
\includegraphics{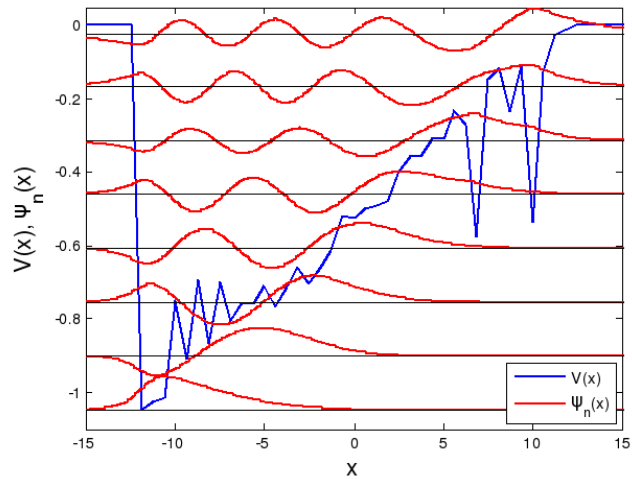}
\caption{Close up view of
Figure~\ref{fig:vxx} showing the part of the domain
  that represents the molecule.}
\label{fig:vxx-close-up}
\end{figure}

\begin{figure}
\includegraphics{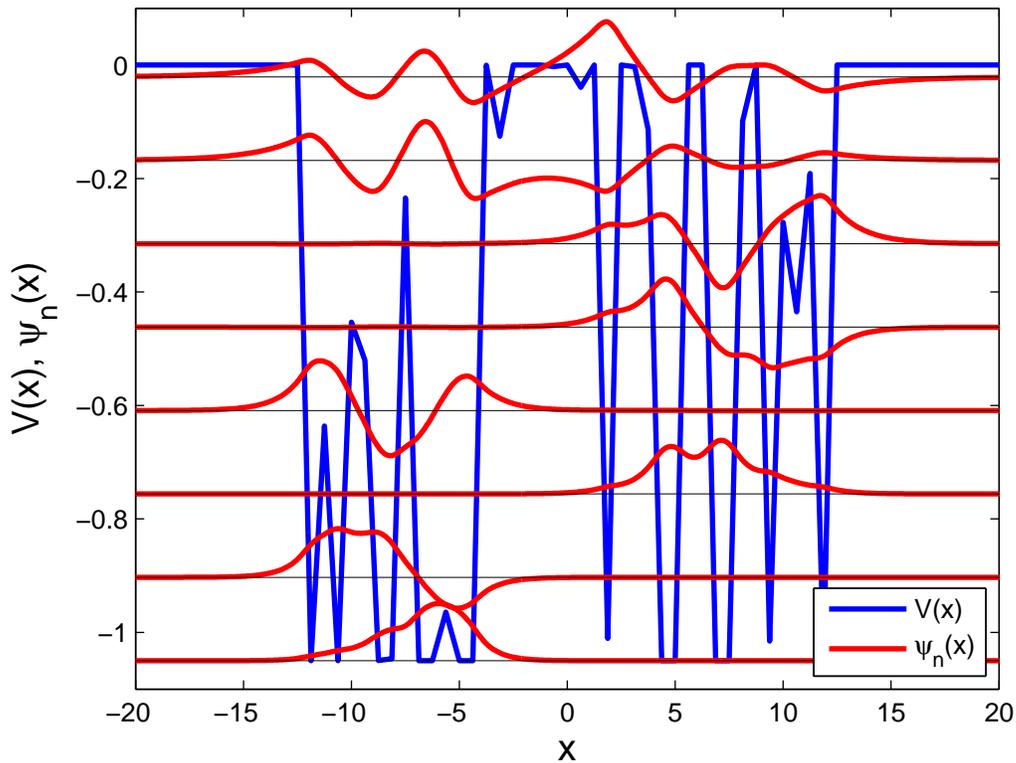}
\caption{Close up view of
Potential energy and wavefunction when the intrinsic hyperpolarizability is optimized.  Starting potential is $x+10 \sin(x)$.}
\label{fig:fig:x+10sin(x)_n20_opt_zoom_bold}
\end{figure}

Figure \ref{fig:fig:x+10sin(x)_n20_opt_zoom_bold} shows the optimized potential energy function for a starting potential of the form $x+10 \sin (x)$.  The potential energy function is characterized by large oscillations and the wavefunctions are bimodally localized near $x = \pm 0.8$.  However, the wavefunctions are not each individually localized in a unique region as was found for the case Studied by Zhou, when the amplitude of wavefunction oscillations was not restricted.

As found in our previous work, the optimized potentials each share certain universal properties.\cite{kuzyk10.01} For example, when the intrinsic hyperpolarizability is optimized, only two excited states dominate the sum-over states expression, that is, two excited states are responsible for over 90\% of the hyperpolarizability.   This is consistent with the three-level ansatz.

Table 1 summarizes the results for the full set of
calculations. This table includes the values of the optimized
hyperpolarizability for all three methods of calculation, as
well as the hyperpolarizability of the starting
potential. The SOS and DF values were calculated after
optimizing $\beta_{NP}$. $\beta_{NP}$ and $\beta_{SOS}$ agree to within about 0.1\%, suggesting that the numerical calculations are accurate.  $\beta_{DF}$ is typically within 5\% of the other two, which is commonly observed when $\beta_{DF}$ and $\beta_{SOS}$ have converged.\cite{kuzyk05.02,kuzyk06.01}

Table 1 shows that all optimized intrinsic hyperpolarizabilities, independent of the starting potential, are around 0.7 and never larger than 0.708.   Also $E_{10}/E_{20}$ is between 0.45 and 0.49, and $x_{10}/x_{10}^{max} \approx 0.79$ for all optimized potentials. This universal behavior is in agreement with previous 1D\cite{zhou06.01,zhou07.02,zhou08.01,kuzyk10.01} calculations, calculations that optimize the positions of nuclei in 2D,\cite{kuzyk06.02} and when the effects of externally-applied electromagnetic fields are included.\cite{watkins09.01}
\begin{widetext}

\begin{table}\scriptsize
\caption{Summary of calculations with different starting
    potentials.  $\beta_s$ is the hyperpolarizability of the starting
    potential while the other ones are after optimization.  The
    transition moments and energies are in dimensionless units.  When one desires $x_{nm}$ to be
    in units of angstroms, then energies would
    be determined by multiplying all values of $E_{n0}$ by
    $\hbar^2/ma^2$, with $a=10^{-10} \, m$ ($1 \,{\AA}$).  In this
    case, the energy is in units of $1.2 \times 10^{-18} \, J$ or
    about $7.6 eV$.\label{tab:Vsumary}}
\begin{tabular}{c c c c c c c c c c c c c c c c}
  \hline
  $V(x)$ & $\beta_{S}$ & $\beta_{SOS}$ & $\beta_{DF}$ & $\beta_{NP}$ &
  $\tau_{00}^{(80)}$ & $E_{10}$ & $E_{20}$ & $x_{00}$ & $x_{10}$ &
  $x_{20}$ & $x_{11}$ & $x_{21}$ & $x_{22}$ &
  $\frac{x_{10}}{x_{max}}$ & $\frac{E_{10}}{E_{20}}$ \\
  \hline
0 & 0 & 0.6859 & 0.6705 & 0.6859 & 0.0122 & 0.020 & 0.045 & -5.880 &
-3.911 & 1.406 & 2.757 & 5.531 & 2.855 & -0.786 & 0.453
\\
$\tanh(x)$ & 0.0507 & 0.7073 & 0.6841 & 0.7071 & 0.011 & 0.100 & 0.208
& -9.660 & 1.760 & -0.602 & -5.663 & -2.561 & -4.414 & 0.789 & 0.484
\\
$x$ & 0.6447 & 0.7081 & 0.6936 & 0.7080 & 0.0069 & 0.119 & 0.247 &
-9.562 & -1.619 & 0.551 & -5.894 & 2.347 & -4.682 & -0.789 & 0.482
\\
$x^{2}$ & 0.5568 & 0.7084 & 0.7054 & 0.7084 & 0.0013 & 0.088 & 0.182 &
-8.501 & -1.881 & 0.643 & -4.227 & 2.734 & -2.871 & -0.789 & 0.483
\\
$\sqrt{x}$ & 0.6650 & 0.7071 & 0.6809 & 0.7068 & 0.0127 & 0.114 &
0.237 & -9.812 & -1.656 & 0.574 & -6.060 & 2.402 & -5.045 & -0.789 &
0.479
\\
$x + \sin x$ & 0.4854 & 0.7078 & 0.6973 & 0.7077 & 0.0052 & 0.115 &
0.240 & -9.498 & -1.645 & 0.561 & -5.776 & 2.383 & -4.557 & -0.790 &
0.481
\\
$x + 10 \sin x$ & 0.2248 & 0.6822 & 0.6875 & 0.6821 & 0.0056 & 0.191 &
0.413 & -6.300 & -1.278 & 0.497 & -9.138 & -1.820 & -8.979 & -0.791 &
0.463 \\
\hline
\end{tabular}
\end{table}
\end{widetext}

Thus, we find that in the more restrictive case where the amplitude of changes in the potential energy function are constrained, and when moving the walls away from the molecule to approximate continuum states, the universal properties that are observed have not changed.  However, while the three-level ansatz continues to be observed, we have not observed full localization of the wavefunctions.  But, we do observe the same sort of oscillations, suggesting that modulation of conjugation may yet prove to be a good paradigm for enforcing the three-level ansatz and resulting in an optimized nonlinear-optical response of real molecules.\cite{perez09.01}

\section{Conclusion}

Many potential energy
functions, even in the more restricted case studied here, are found to bring the intrinsic hyperpolarizability close to the
fundamental limit. In particular, there appear to be two classes of
potentials that approach this limit. First, the wiggly potential
energy functions are found to have only somewhat spatially separated
eigenfunctions.  These potentials led to the prediction that modulation
of conjugation may show promise for higher values of the intrinsic
hyperpolarizability. The second class was characterized by much
smaller wiggles. Though the shapes of these potentials varied
significantly, they were found to have several features in common. In
particular, upon optimization, $\beta_{int}$ approached
0.71, $E_{10}/E_{20}$ is between
0.45 and 0.49 and $x_{10}/x_{10}^{max} \approx 0.79$.
Also, when the hyperpolarizability is
optimized the system is dominated by three states, so the three-level
ansatz holds.

It is interesting that so many differently shaped
potentials end up having so many similar characteristics and that they share certain universal properties.\cite{kuzyk10.01}  This hints at the possibility for new underlying physics.  Since there
appear to be a large number of different potential energy functions
that lead to a maximized intrinsic hyperpolarizability, it may be
possible to use this fact to engineer molecules that achieve ever larger
intrinsic hyperpolarizabilities that approaches the fundamental.

{\bf Acknowledgements: } MGK thanks the National Science Foundation (ECCS-0756936) for supporting this work.

%\bibliography{\bibs}

\clearpage

\end{document}